\documentclass[12pt,square,numbers]{elsarticle}
\usepackage[hmargin=1in,vmargin=1in]{geometry}
\usepackage{graphicx} 
\usepackage{float}
\usepackage{multirow}
\usepackage{array}
\usepackage{setspace}

\usepackage[normalem]{ulem}
\usepackage{enumerate}
\usepackage{enumitem}
\usepackage{xcolor}
\usepackage{tabularx}
\usepackage{hyperref}
\usepackage{natbib}

\begin{document}

\title{Revisiting Safe Temperature for Environmental Accelerated Aging of Additively Manufactured Polymers}

\author[main,main2]{Kevin LoGiudice\fnref{equal}}
\author[main]{Keven Alkhoury\fnref{equal}}
\author[main2]{Nikash Long}
\author[main]{Justin Moustouka}
\author[main]{Nour Mousbah}
\author[main3]{Irine Chenwi}
\author[main3]{James LeBlanc}
\author[main,main2]{Vikas Srivastava\corref{cor1}}
\ead{vikas\_srivastava@brown.edu}
\cortext[cor1]{Corresponding author}
\fntext[equal]{These authors contributed equally to this work.}
\address[main]{School of Engineering, Brown University, 184 Hope Street, Providence, RI 02906 USA}
\address[main2]{Institute for Biology, Engineering and Medicine, Brown University, Providence, RI 02912, USA}
\address[main3]{Naval Undersea Warfare Center (Division Newport),
1176 Howell St, Newport, RI 02841 USA}

\begin{abstract}
Accelerated aging is widely used to study the long-term behavior of materials within laboratory time scales, particularly for materials exposed to solvent environments over extended periods. This is
especially important for additively manufactured (AM) polymers, whose increasing use in naval and commercial undersea applications requires reliable methodologies for assessing durability under in-service conditions.
A common approach relies on elevating the temperature below the glass transition or melting temperature to accelerate degradation. However, the temperature limits for accelerated aging of AM polymers remain poorly understood, particularly because temperatures beyond a threshold may activate deformation and degradation mechanisms that are absent under service conditions. To address this gap, this paper investigates fused deposition modeling (FDM) Acrylonitrile Butadiene Styrene (ABS) exposed to saltwater and deionized (DI) water to establish a temperature threshold for accelerated aging in aqueous environments and propose a methodology for determining such thresholds. Controlled geometries and varying print directions were employed to explicitly probe the underlying mechanisms. We show that samples exposed to temperatures above the threshold exhibit pronounced shrinkage and warping along the printing direction due to the relaxation of process-induced internal stresses. These observations establish an accelerated-aging temperature threshold of 50$^\circ$C for ABS, beyond which additional mechanisms absent under service conditions become active. Additionally, the resulting geometric distortions are masked in thick geometries but become highly pronounced in thin structures. Moreover, solvent ionic content strongly influences water uptake, with saltwater reaching saturation within 1 day, whereas DI water did not reach saturation even after 30 days and exhibited significantly greater mass uptake.

\bigskip 


\end{abstract}

\maketitle

\section{Introduction}
The commercialization of polymers, both thermoplastics and thermosets, marked a shift in engineering materials, offering a multifunctional alternative to traditional materials such as metals, wood, glass, and ceramics. Their durability, high strength-to-weight ratio, low cost, chemical resistance, and manufacturing versatility have enabled widespread use across applications ranging from packaging and textiles to structural components and biomedical technologies \citep{edlund2002degradable,siracusa2008biodegradable,srivastava2010thermally,silvestre2011food,ulery2011biomedical,ozdil2014polymers,guo2018conducting,mangaraj2019application,Kazemi2022state}. The reach of these applications is equally facilitated by advanced manufacturing methods, such as additive manufacturing (AM) or 3D printing.\footnote{We use the terms additive manufacturing and 3D printing interchangeably. While ``additive manufacturing'' is the preferred term within the scientific and engineering community, ``3D printing'' is also widely used to describe the layer-by-layer fabrication of components from a digital model.} AM encompasses a variety of methodologies, including fused deposition modeling (FDM), stereolithography (SLA), digital light processing (DLP), selective laser sintering (SLS), multi-jet fusion (MJF), and more newly emerging techniques \citep{Srinivasan2021AM}. While various AM techniques are equally important, this work focuses on FDM due to its accessibility and widespread adoption across many sectors \citep{wickramasinghe2020fdm,kristiawan2021review,ma2026polymer,gupta2026TPMSfracture}.

A wide range of semi-crystalline and amorphous thermoplastic materials can be printed via FDM \citep{rahim2019recent}. We use ABS in this work for its versatility across biomedical, structural, and military applications \citep{mccullough2013surface,olivera2016plating,mallikarjuna2025review}. Since materials are routinely subjected to environmental stressors during service, such as elevated temperatures, pressure, UV radiation, humidity, and water, they are susceptible to degradation via various chemical reactions \citep{pasparakis2012photodegradable,delplace2015degradable}. For naval maritime and commercial subsea applications, where polymer composites and polymers are continuously exposed to aqueous and saline environments, local interfacial damage for composites and hydrolysis for bulk polymers become dominant degradation mechanisms \citep{elsawy2017hydrolytic,leblanc2023effect,leblanc2024high,Vaishakh2024,chen2025hydrolytic}. In hydrolysis of polymers, water molecules induce chain scission, progressively reducing the polymer’s molecular weight and ultimately compromising its mechanical integrity over time.

Characterizing the in-service performance of AM polymers under environmental stressors is essential, as degradation often governs premature failure. However, real-world aging occurs over timescales that far exceed those accessible in laboratory settings, which requires the development of accelerated aging methodologies to simulate service conditions within laboratory time frames \citep{frigione2021can}. Accelerated aging typically involves exposing materials to conditions more severe than those encountered in service to reproduce the effects of long-term exposure within a shorter duration. However, care must be taken to ensure that these conditions remain representative and do not introduce degradation mechanisms that would not occur under realistic operating environments. Among the available approaches, temperature has been widely used to accelerate polymer aging, often motivated by time-temperature superposition (TTS). It is critical to remain within appropriate time/temperature limits to avoid altering the underlying physics \citep{matsumoto1988time}. 

In this work, we examine the upper temperature bounds for accelerated aging to identify conditions that enable meaningful acceleration without altering the underlying physics. Through carefully designed experiments using controlled geometries and varying printing directions, we show that elevated temperatures alone can induce significant shrinkage and warping, likely associated with the relaxation of process-induced internal stresses in the 3D-printed polymer during extrusion, resulting in partial recovery toward its initial configuration, with these effects becoming more pronounced in the presence of solvent. Such effects can alter the degradation mechanisms, indicating that increasing temperature may not only accelerate degradation but also change its mechanisms, even when operating below the material's glass transition and/or melting temperatures. 

The novelty of this work lies in demonstrating that temperature cannot always be treated as an acceleration parameter for polymer aging across a wide range, as increasing temperature beyond a material-dependent limit could activate mechanisms that are not representative of in-service conditions. The results show that a valid accelerated aging protocol is governed by the coupled effects of temperature, geometry, and 3D printing direction. We also show that thermally activated deformation mechanisms, such as shrinking and warping, become more pronounced in thin or geometrically vulnerable FDM-printed structures, while the printing direction can strongly influence whether these instabilities are activated or even observed. These behaviors are likely associated with the printed architecture and process-induced residual stresses, rather than representing intrinsic changes in the bulk ABS material. These findings further motivate exploring alternative accelerated aging strategies beyond simply increasing temperature, such as modifying salt concentration, hydrostatic pressure, or other environmental driving forces, that could accelerate degradation while preserving the underlying in-service physics.

\section{Materials and Methods}

\subsection{Material fabrication}
ABS filament (1621, Ultimaker) was stored in a humidity-controlled Material Station Manager (218053, Ultimaker) controlled at 25\% relative humidity, and specimens were fabricated using an Ultimaker S8 3D printer via extrusion through an AA+ print core with a 0.4 mm nozzle diameter (237391, Ultimaker). The specimens were printed with an extruder temperature of 250$^\circ$C and a print bed temperature of 85$^\circ$C. The nominal line width was set to 0.4 mm, with a layer height of 0.2 mm and a print speed of 150 mm/s. Specimen prints consisted of 100\% infill lines, with no walls or top/bottom layers. A linear infill pattern was employed, consisting of discontinuous, parallel lines at varying raster angles. The layers were printed along the specimen height.

\begin{figure}[H]
    \centering
\includegraphics[width=.85\textwidth]{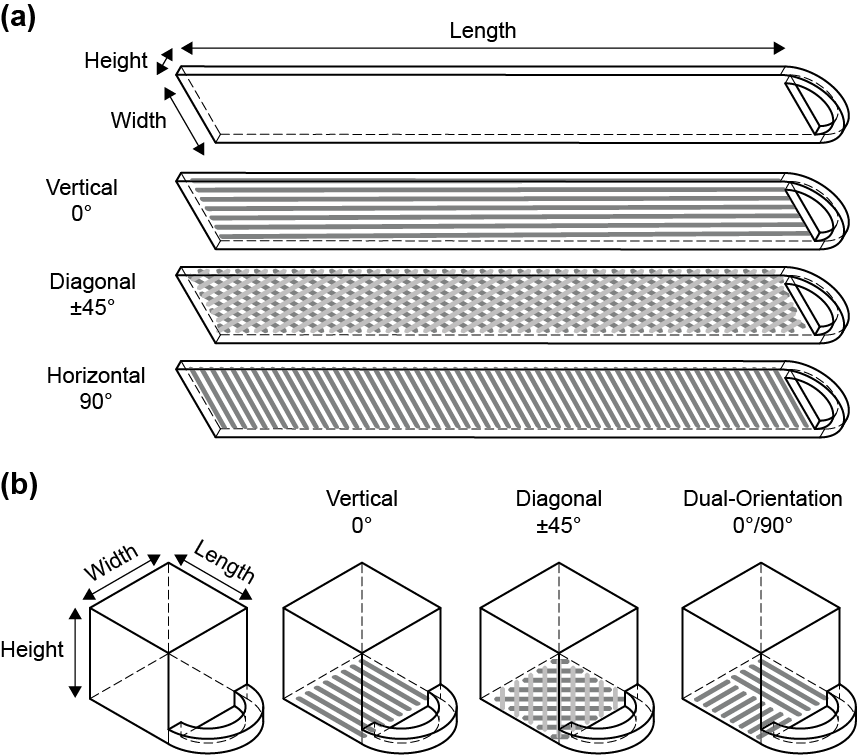}	\caption{Schematic of the representative geometries and print orientations considered in this study, including (a) a rectangular specimen with dimensions 100  mm $\times$ 13 mm $\times$ 1.6 mm ($L \times W \times H$), and (b) a cubic specimen with dimensions 10 mm $\times$ 10 mm $\times$ 10 mm. Note that the layers were printed along the specimen height.}
\label{specimen_geometries}
\end{figure}

Two distinct geometries, rectangles and cubes (see Figure \ref{specimen_geometries}), were used to investigate the temperature limit threshold for accelerated aging by considering the effects of (i) temperature, (ii) solvent, and (iii) their coupling. 
The raster angle was varied to achieve distinct line orientations for each geometry type. Rectangle specimens were oriented as either vertical print lines (print lines oriented along the longitudinal axis, 0$^\circ$ to the long axis), horizontal print lines (print lines oriented along the width, 90$^\circ$ to the long axis), or diagonal print lines alternating $\pm45^\circ$ to the long axis by layer. Cube specimens were oriented as either vertical print lines (print lines oriented along one cube face direction, 0$^\circ$), diagonal print lines alternating $\pm45^\circ$ by layer, or dual-orientation print lines (half the print lines oriented along 0$^\circ$ and half along 90$^\circ$ within each layer). Horizontal print lines (90$^\circ$) were not considered separately because they are equivalent to the vertical orientation by symmetry. Each geometry served a complementary purpose. The rectangular specimens enabled clear qualitative observation of dimensional changes; however, significant out-of-plane warping limited their use for reliable quantitative measurements. To address this limitation, cubic specimens were introduced, enabling more accurate quantification of dimensional changes under identical conditions.\footnote{While the cubic specimens are capable of providing both qualitative and quantitative measurements, the thinner rectangular specimens were retained because their geometry proved more sensitive to temperature-induced instabilities, revealing warpage and shrinkage mechanisms that were not observed in the thicker cubic specimens under certain exposure conditions.} For both geometries, mass uptake was measured throughout the study. Additionally, all specimens were designed with integrated mounting hooks to ensure they remained freely suspended and uniformly exposed to their surroundings (temperature and/or solvent) throughout the experiments, without boundary constraints, thereby preventing stress coupling due to mechanical constraints.\footnote{The use of hooks may appear trivial; however, they were necessary to allow the specimens to deform freely during exposure. Preliminary experiments in which specimens were placed flat on a surface at elevated temperature showed that contact (friction) with the surface inhibited bending and prevented us from observing the true deformation behavior.}

\subsection{Exposure conditions and protocol}
\begin{figure}[H]
    \centering
\includegraphics[width=.85\textwidth]{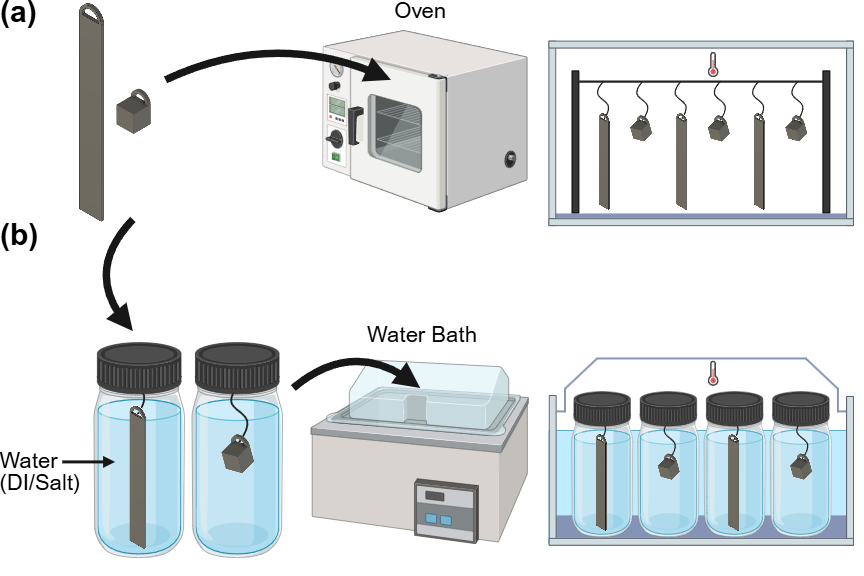}
\caption{Schematic of the experimental setup. A rectangular or cubic specimen is attached by the mounting hook, 3D-printed into the geometry, to hang freely in the selected environment. (a) Specimens kept in air were temperature-controlled in a vacuum oven. (b) Specimens exposed to deionized water or saltwater were placed into a secondary container of the solvent, which was then placed into a temperature-controlled water bath. Graphic generated via \citet{BioRender_Product_2026}.}
\label{fig:methods_illustration}
\end{figure}

To identify elevated temperatures suitable for accelerated aging, specimens (rectangular and cubic) were exposed to temperatures selected relative to the glass transition temperature of ABS (Tg = 100.5$^\circ$C, from the supplier's technical datasheet \citep{Ultimaker_ABS_TDS_2022}): 95$^\circ$C, 65$^\circ$C, 50$^\circ$C, and 22$^\circ$C (room temperature).\footnote{Although the glass transition temperature, Tg, can decrease as the polymer degrades, most of the phenomena observed in the present study occurred within the first 24 hours of exposure, which is considerably shorter than the timescales associated with thermal degradation. Moreover, although solvent swelling can reduce Tg of polymers \citep{pan2026constitutive}, similar shape changes were observed even under dry conditions. For these reasons, Tg was assumed not to have changed significantly in this work.} Specifically, since our aim was to isolate and examine the effects of (i) temperature, (ii) solvent, and (iii) their coupling, specimens were exposed to these temperatures either in air, in 3.5\% saltwater solution according to ASTM D1141 standard \citep{ASTM_D1141_2021}, or in deionized (DI) water according to ASTM D1193 Type II \citep{ASTM_D1193_2024}. The inclusion of both deionized (DI) water and saltwater exposures provided crucial insight that will be discussed later. For air exposures, samples were freely suspended from a hook and placed in a vacuum oven at ambient pressure (3618-5, Thermo Scientific), as schematically shown in Figure \ref{fig:methods_illustration}a, ensuring they remained unconstrained throughout the exposure. For solvent environments, specimens were freely immersed in the selected solution in sealed glass containers, which were then placed in a temperature-controlled water bath (FSGPD20, Fisherbrand) at the prescribed setpoints, as schematically shown in Figure \ref{fig:methods_illustration}b. Over the course of the experiments, the fluid bath level was maintained by periodically adding small amounts of water to compensate for evaporation while ensuring that temperature perturbations were negligible.  Solvent exposures at room temperature were performed in glass containers on a lab benchtop in a room maintained at approximately 22 $^\circ$C.
Measurements were taken at 2, 4, 8, 12, and 24 hours to capture the faster initial diffusion, then daily up to 7 days, followed by 4-day intervals thereafter up to 30 days. At each time point, three specimens were tested to ensure repeatability. Specifically, images of the rectangular specimens were captured immediately after long-term exposure using a digital camera (9338B001, Canon) mounted on a tripod to qualitatively monitor dimensional changes induced by temperature, solvent diffusion, and their coupling. For cubic specimens, dimensional changes in length, width, and height were measured immediately after long-term solvent exposure using vernier calipers (DCLA-0605, Vinca). In addition, the mass of all specimens was recorded using a precision balance (XS64, Mettler Toledo) with an accuracy of 0.1 mg. Since the smallest specimen mass was approximately 1000 mg, the measurements provided sufficient resolution to detect any measurable water uptake. 
The dimensional measurements, image acquisition, and mass measurements were performed with the specimens temporarily removed from their respective exposure conditions for approximately one minute in total, after which they were immediately returned.

\section{Results and Discussion}

\subsection{Air exposure}

\begin{figure}[H]
    \centering	\includegraphics[width=.85\textwidth]{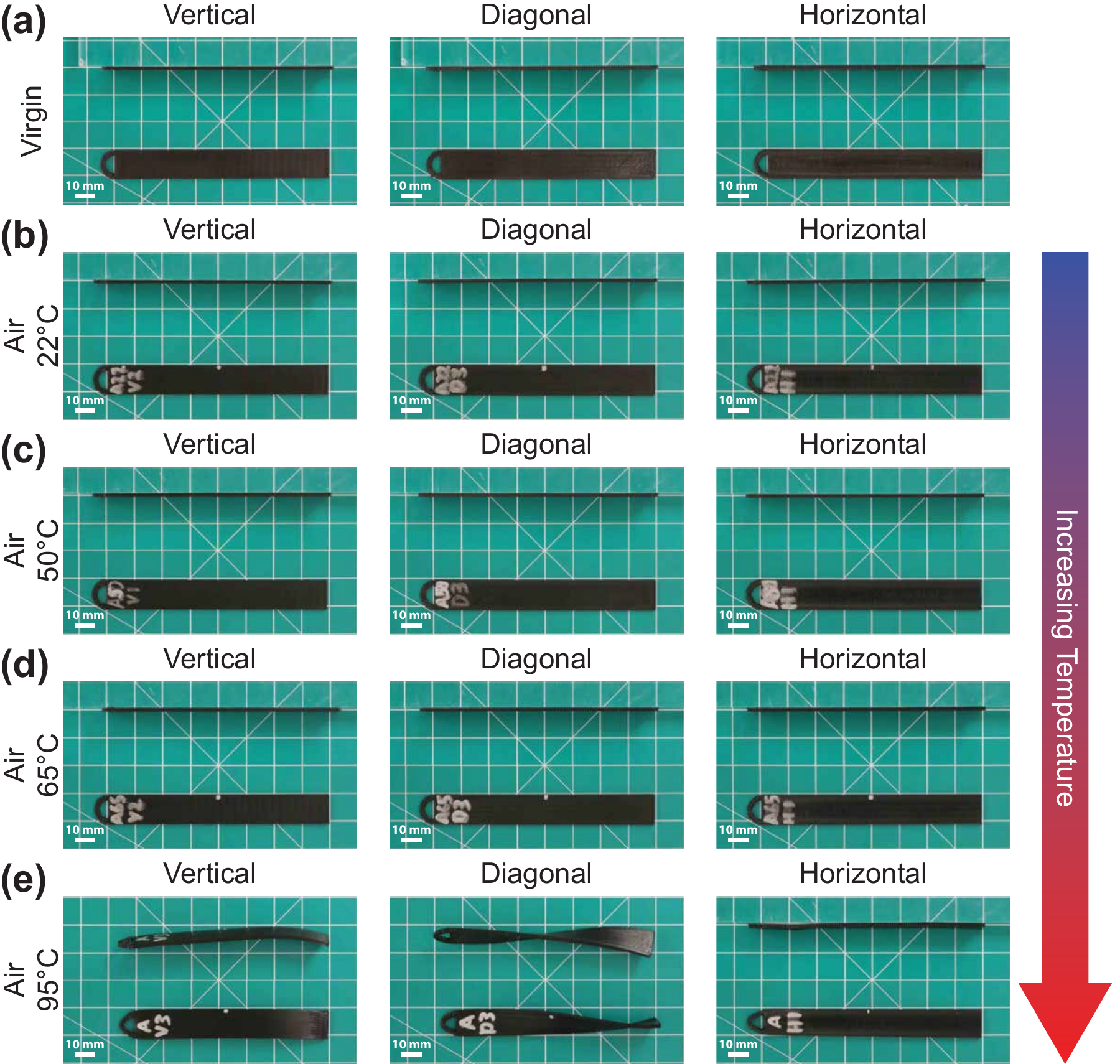}
	\caption{Qualitative images of rectangular specimens showing the shrinkage and out-of-plane warping for (a) virgin samples; specimens exposed to air for 30 days at (b) 22$^\circ$C, (c) 50$^\circ$C, (d) 65$^\circ$C, and (e) 95$^\circ$C. Note that a support was used in the side-view images to prevent the specimens from falling during imaging. }
	\label{fig:AirRectangles}
\end{figure}

At 95$^\circ$C, qualitative dimensional changes were observed after just 2 hours of exposure and became more pronounced over a 30-day exposure period. Rectangular samples exhibited irreversible shrinkage in the print direction, which in turn resulted in an out-of-plane warpage following the pattern shown in Figure \ref{fig:AirRectangles}. The vertical samples exhibited shrinkage along their length (the printing direction), resulting in pronounced out-of-plane warping. The diagonal samples exhibited a twisting deformation pattern, attributed to shrinkage along the diagonal print lines that alternated at $\pm45^\circ$. The horizontal samples primarily shrank along their width (the printing direction) while exhibiting a slight increase in length. No mass change was detected.
On the other hand, cubic samples showed no measurable changes in length, width, or height. No mass change was detected.
This suggests that the elevated temperature of 95$^\circ$C activated an irreversible shrinkage mechanism, likely associated with the relaxation of process-induced internal stresses in the 3D-printed polymer during extrusion, resulting in partial recovery toward its initial configuration. This behavior may be interpreted as a shape memory-like response. The effect appears more pronounced in thinner structures and is not evident in thicker (bulk) cubic specimens. This was further investigated by examining the coupling between solvent and temperature. We note that no qualitative or quantitative changes were observed by air exposure at 65$^\circ$C, 50$^\circ$C, or 22$^\circ$C.

\subsection{Solvent exposure}
\begin{figure}[H]
    \centering	\includegraphics[width=.85\textwidth]{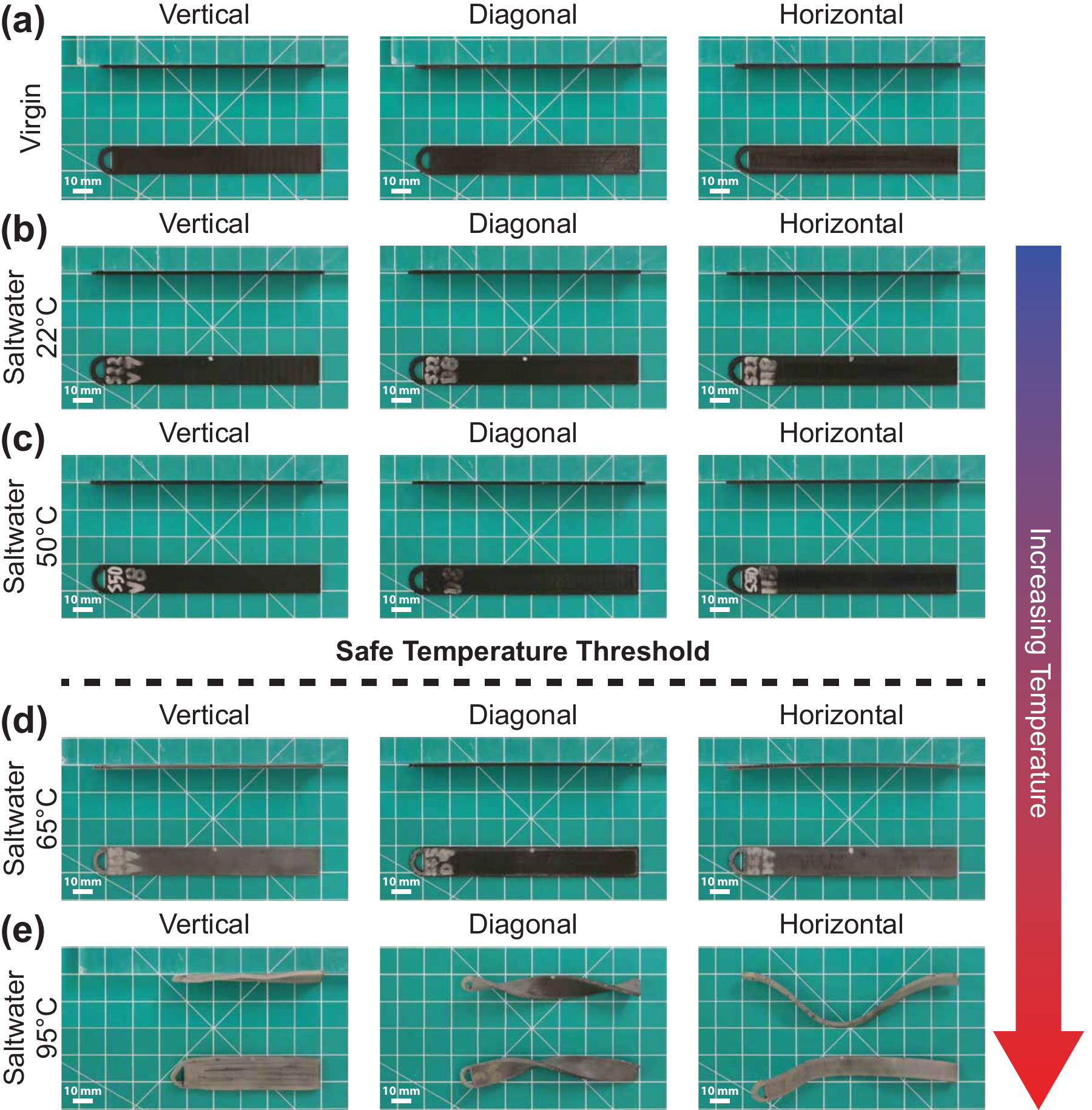}
	\caption{Qualitative images of rectangular specimens showing the shrinkage and out-of-plane warping for (a) virgin samples; specimens exposed to saltwater for 30 days at (b) 22$^\circ$C, (c) 50$^\circ$C, (d) 65$^\circ$C, and (e) 95$^\circ$C. Note that a support was used in the side-view images to prevent the specimens from falling during imaging.}
	\label{fig:SaltwaterRectangles}
\end{figure}

\begin{figure}[H]
    \centering	\includegraphics[width=.85\textwidth]{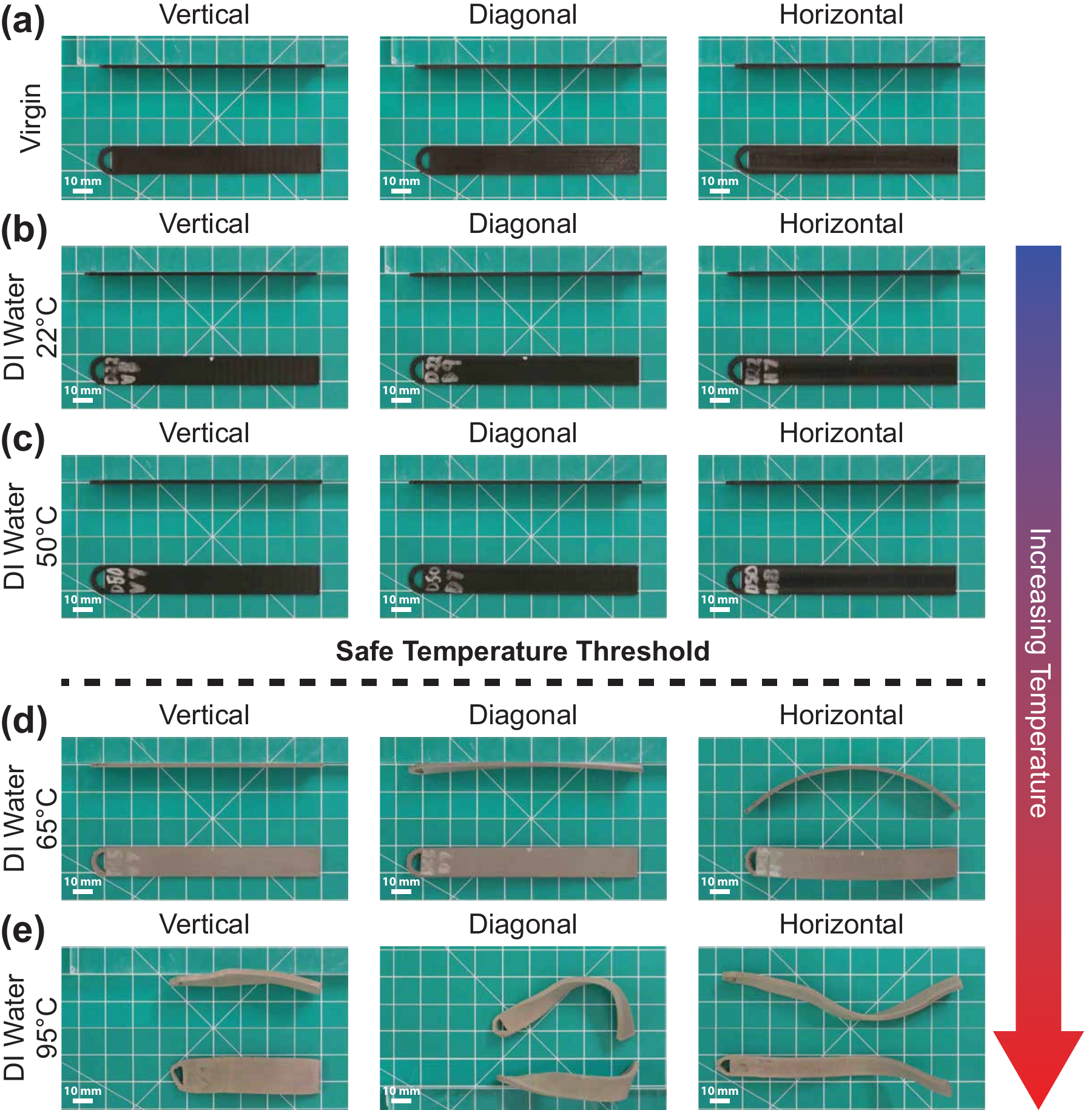}
	\caption{Qualitative images of rectangular specimens showing the shrinkage and out-of-plane warping for (a) virgin samples; specimens exposed to DI water for 30 days at (b) 22$^\circ$C, (c) 50$^\circ$C, (d) 65$^\circ$C, and (e) 95$^\circ$C.  Note that a support was used in the side-view images to prevent the specimens from falling during imaging.}
\label{fig:DIwaterRectangles}
\end{figure}

For solvent exposure at 95$^\circ$C, the rectangular specimens exhibited significant qualitative changes after just 2 hours, with these changes continuing to evolve over a 30-day period. In both saltwater and DI conditions, pronounced shrinkage was observed along the print direction, resulting in notable out-of-plane warpage patterns, as shown in Figures \ref{fig:SaltwaterRectangles} and \ref{fig:DIwaterRectangles}. The vertical samples exhibited shrinkage along their length (the printing direction), resulting in shortening and pronounced out-of-plane warping. The diagonal samples exhibited a twisting deformation pattern, attributed to shrinkage along the diagonal print lines that alternated at $\pm45^\circ$. The horizontal samples primarily shrank along their width (the printing direction) while expanding more pronouncedly in the transverse (length) direction. These effects were more pronounced in DI water than in saltwater due to higher water uptake, although the overall deformation trends remained consistent between the two environments.

\begin{figure}[H]
    \centering
\includegraphics[width=.85\textwidth]{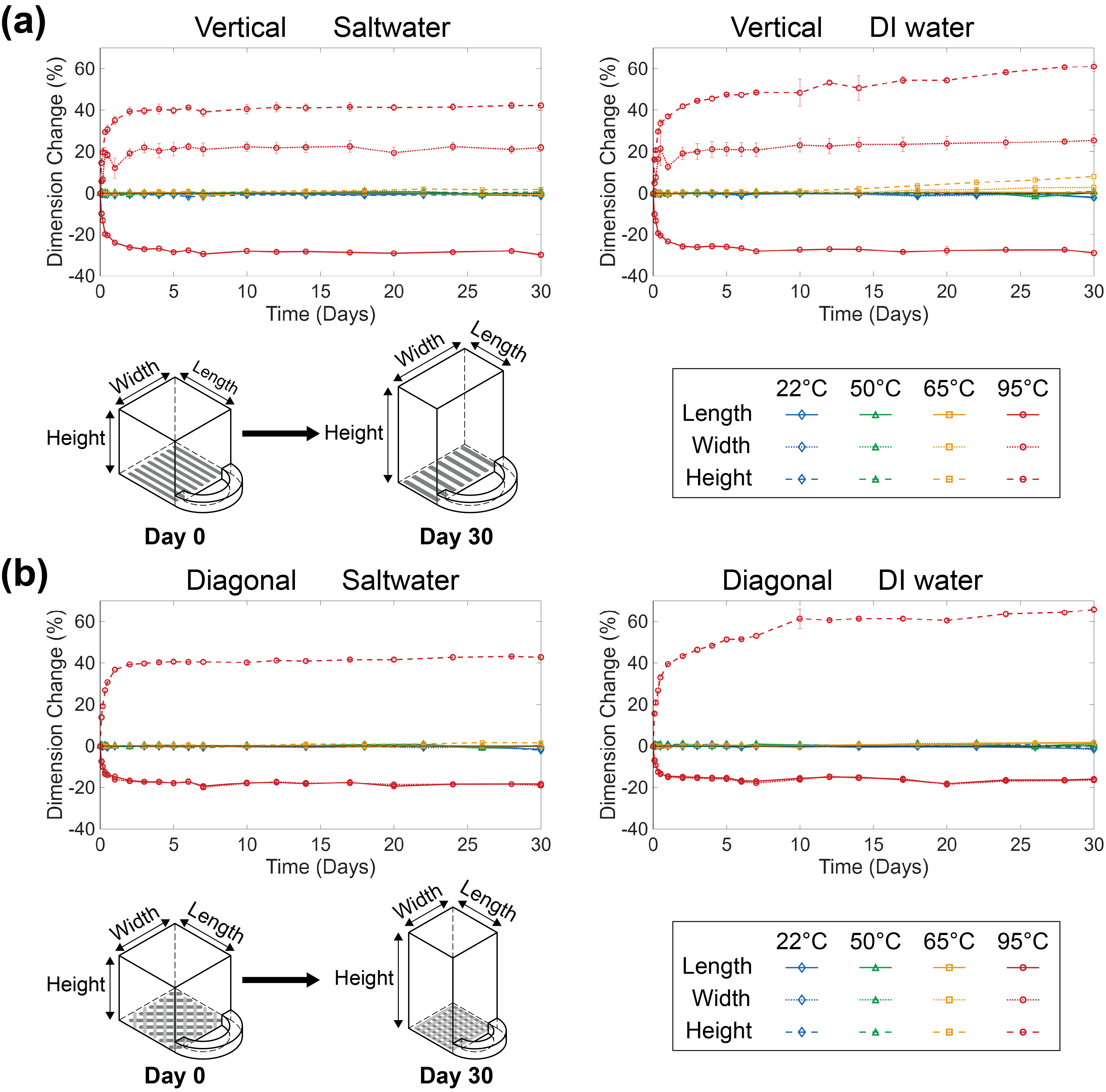}
	\caption{Dimensional evolution of cubic specimens exposed to saltwater and DI water at 22$^\circ$C, 50$^\circ$C, 65$^\circ$C, and 95$^\circ$C, along with schematics of the reference and deformed cube geometries for specimens of (a) vertical line orientation, and (b) diagonal line orientation. Note that the error bars represent the standard deviation.}
	\label{fig:dimensions}
\end{figure}

The cubic specimens, subjected to the same conditions, exhibited similar qualitative behavior and enabled quantitative characterization, as shown in Figure \ref{fig:dimensions}. Dimension change (\%) is defined as $ \frac{\Delta d}{d_0} = \frac{d_t - d_0}{d_0} \times 100$, where $d_t$ is the dimension (i.e., length, width, or height) at the time of measurement and $d_0$ is the virgin (dry) dimension. 

\begin{figure}[H]
    \centering
\includegraphics[width=.85\textwidth]{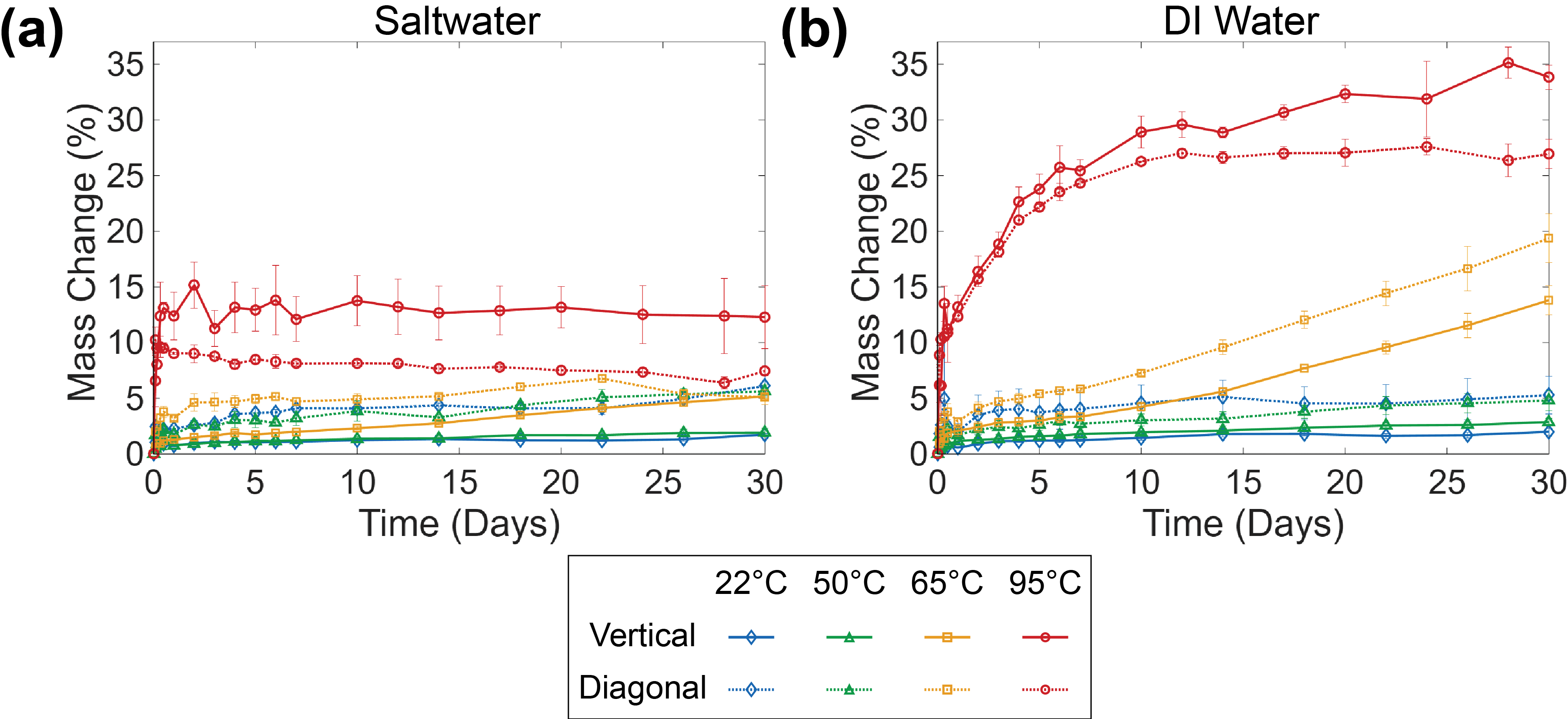}
\caption{Mass uptake profiles over 30 days for cubic specimens exposed to 22$^\circ$C, 50$^\circ$C, 65$^\circ$C, and 95$^\circ$C in (a) saltwater, and (b) DI water. Note that the observed differences between specimen orientations suggest potential orientation-dependent behavior, which is not investigated further here and remains outside the scope of the present study. Also note that the error bars represent the standard deviation.}
	\label{fig:mass}
\end{figure}

In both saltwater and DI environments, shrinkage along the print direction and the associated dimensional changes were consistently observed across orientations (vertical and diagonal). However, the effect was more pronounced in DI water compared to saltwater. The mass change (\%) is defined as $\frac{\Delta m}{m_0} = \frac{m_t - m_0}{m_0} \times 100$, where $m_t$ is the mass at the time of measurement, and $m_0$ is the virgin (dry) mass. In saltwater, specimen saturation was reached, and the corresponding dimensional changes stabilized as equilibrium was approached, as shown in Figure \ref{fig:mass}a. In contrast, DI exposure did not exhibit mass saturation over the 30-day exposure period, as seen in Figure \ref{fig:mass}b, and the dimensional changes continued to evolve.\footnote {The slower, non-saturating uptake in DI water, compared with the apparent saturation in saltwater, suggests that solvent uptake is not governed solely by the water concentration gradient and may depend on solvent-specific interactions among the polymer, water, and dissolved ions.}
The mass and dimensional changes of cubes printed in the vertical and diagonal orientations after 5 and 30 days of exposure at 95$^\circ$C are summarized in Table \ref{tab:MassDimensions}.
\begin{table}[H]
\centering
\caption{Mass and dimensional changes of cubes printed with (a) vertical and (b) diagonal 3D print orientations, exposed at 95$^\circ$C.} 
\label{tab:MassDimensions}
\textbf{(a) Vertical orientation}
\vspace{0.5em}
\renewcommand{\tabularxcolumn}[1]{m{#1}}
\renewcommand{\arraystretch}{1.4}
\begin{tabularx}{\textwidth}{ | l | *{8}{>{\centering\arraybackslash}X |} } \hline
 & \multicolumn{2}{c|}{\textbf{$\frac{\Delta m}{m_0}$}} 
 & \multicolumn{2}{c|}{\textbf{$\frac{\Delta l}{l_0}$}} 
 & \multicolumn{2}{c|}{\textbf{$\frac{\Delta w}{w_0}$}} 
 & \multicolumn{2}{c|}{\textbf{$\frac{\Delta h}{h_0}$}} \\
 \cline{2-9}
 \textbf{Exposure} & \textbf{5} \par \textbf{days} & \textbf{30} \par \textbf{days} & \textbf{5} \par \textbf{days} & \textbf{30} \par \textbf{days} & \textbf{5} \par \textbf{days} & \textbf{30} \par \textbf{days} & \textbf{5} \par \textbf{days} & \textbf{30} \par \textbf{days} \\
 \hline
 Saltwater & 12.9\%  & 12.3\%  & -28.5\%  & -29.8\%  & 21.3\%  & 21.9\%  & 39.8\%  & 2.4\%  \\
 \hline
 DI Water  & 23.8\%  & 33.8\%  & -25.9\%  & -28.9\%  & 21.1\%  & 25.4\%  & 47.4\%  & 61.1\%  \\
 \hline
 Air       & 0.0\% & 0.0\% & -0.9\% & -0.6\% & -0.6\% & -0.8\% & -0.1\%  & -0.2\% \\
 \hline
\end{tabularx}
\end{table}
\begin{table}[H]
\centering
\textbf{(b) Diagonal orientation}
\vspace{0.5em}
\renewcommand{\tabularxcolumn}[1]{m{#1}}
\renewcommand{\arraystretch}{1.4}
\begin{tabularx}{\textwidth}{ | l | *{8}{>{\centering\arraybackslash}X |} } \hline
 & \multicolumn{2}{c|}{\textbf{$\frac{\Delta m}{m_0}$}} 
 & \multicolumn{2}{c|}{\textbf{$\frac{\Delta l}{l_0}$}} 
 & \multicolumn{2}{c|}{\textbf{$\frac{\Delta w}{w_0}$}} 
 & \multicolumn{2}{c|}{\textbf{$\frac{\Delta h}{h_0}$}} \\
 \cline{2-9}
 \textbf{Exposure} & \textbf{5} \par \textbf{days} & \textbf{30} \par \textbf{days} & \textbf{5} \par \textbf{days} & \textbf{30} \par \textbf{days} & \textbf{5} \par \textbf{days} & \textbf{30} \par \textbf{days} & \textbf{5} \par \textbf{days} & \textbf{30} \par \textbf{days} \\
 \hline
 Saltwater & 8.5\%  & 7.4\%  & -17.9\%  & -18.3\%  & -17.7\%  & -19.0\%  & 40.6\%  & 42.7\%  \\
 \hline
 DI Water  & 22.2\%  & 27.0\%  & -15.4\%  & -15.9\%  & -15.9\%  & -16.4\%  & 51.4\%  & 65.8\%  \\
 \hline
 Air       & -0.1\% & 0.0\% & -0.4\% & -0.3\% & -0.6\% & -0.5\% & 0.0\%  & -0.2\% \\
 \hline
\end{tabularx}
\end{table}

\begin{figure}[H]
    \centering
\includegraphics[width=.85\textwidth]{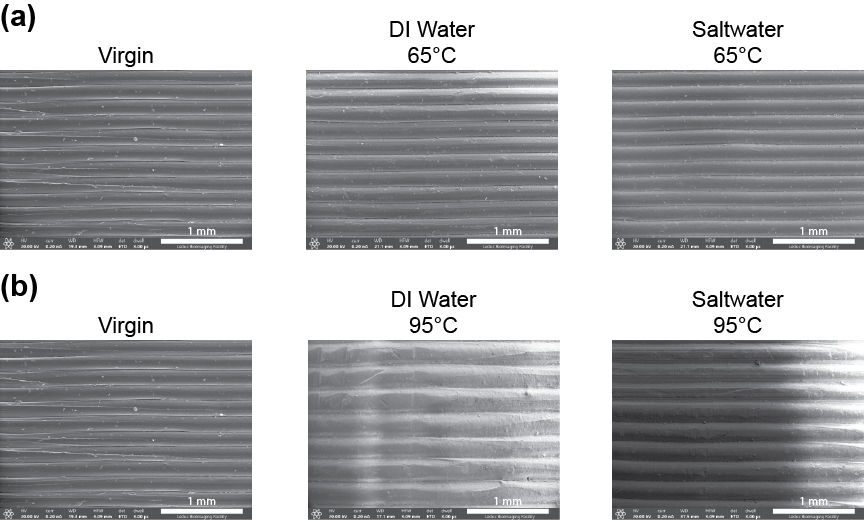}
	\caption{SEM images of virgin, DI water, and saltwater exposed specimens with vertical print orientation at (a) 65$^\circ$C and (b) 95$^\circ$C. Layer heights were measured using ImageJ software \citep{schneider2012nih} and found to be 185.7 $\mu m$ for virgin, 184.3 $\mu m$  for 65$^\circ$C saltwater, 198.8 $\mu m$ for 65$^\circ$C DI water, 240.5 $\mu m$ for 95$^\circ$C saltwater, and 286.7 $\mu m$ for 95$^\circ$C DI water.}
	\label{fig:SEM_Vertical}
\end{figure}

\begin{figure}[H]
    \centering
\includegraphics[width=.85\textwidth]{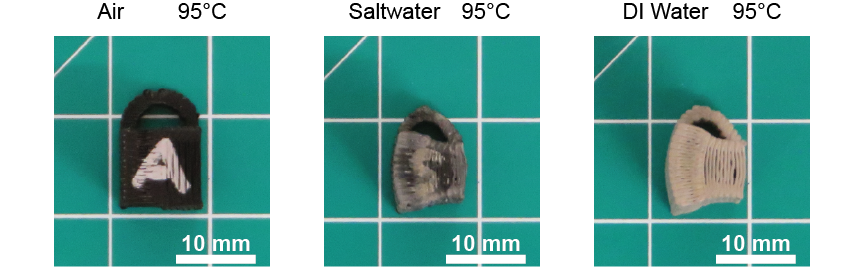}
\caption{Qualitative images of dual-orientation cubic specimens showing the shrinkage in the print direction at 95$^\circ$C in saltwater and DI. We note that although color changes were observed in both saltwater and DI water conditions, with the changes appearing earlier and being more pronounced in DI water, the underlying cause of this difference is beyond the scope of the present work.}
\label{fig:Dual-orientation}
\end{figure}

At 65$^\circ$C, under saltwater exposure, the rectangular specimens largely retained their original geometry except for minor deviations in the horizontally printed specimens. The cubic specimens showed no measurable shrinkage in length, width, or height.\footnote{Although minor changes in height were observed, they were considered negligible after scanning electron microscope (SEM) imaging (see Figure \ref{fig:SEM_Vertical}).} Under DI water exposure, qualitative dimensional changes were observed in the rectangular specimens, particularly in the diagonal and horizontally printed orientations. The cubic specimens also showed minor dimensional changes in height, which were confirmed by SEM imaging (see Figure \ref{fig:SEM_Vertical}), but were less pronounced than those observed at 95$^\circ$C.\footnote{Although a slight increase in height was observed, its origin remains unclear because the specimen width and length remained unchanged. In addition, no evidence was found to support the hypothesis that delamination between layers was responsible for the observed behavior.}
Moreover, exposure to DI water at 65$^\circ$C resulted in a maximum mass increase of approximately 20\%, compared to only 5\% for saltwater exposure. Notably, this 20\% mass increase also exceeds the maximum mass increase observed under saltwater exposure at 95$^\circ$C, which was approximately 15\%. This comparison suggests that solvent uptake primarily contributes to swelling, as evidenced by the slight increase in height, but is not the major driver of the observed shrinkage in length and width (i.e., the shrinkage in length and width is not driven by anisotropic swelling due to solvent uptake). Instead, the shrinkage appears to be primarily temperature-driven, yet solvent-enhanced.
Lastly, the dual-orientation cubic samples shown in Figure \ref{fig:Dual-orientation} provide a clear visualization of the material shrinkage occurring along the printing direction. This again supports our hypothesis: the shrinkage is likely due to the relaxation of process-induced internal stresses in the 3D-printed polymer, leading to a partial recovery toward the initial manufactured configuration, resembling a shape memory-like response.

What is more important is that such behavior is not necessarily representative of service conditions and may instead be an artifact of elevated temperature, even when the exposure temperature remains below the polymer's glass transition or melting temperature. Furthermore, the effect appears to be significantly more pronounced in thin structures, such as rectangular specimens, while remaining largely absent in thick cubic specimens, where the lack of measurable macroscopic deformation could easily lead to the phenomenon being overlooked altogether. This observation suggests that the assessment of accelerated aging protocols should not rely solely on thicker (bulk) geometries but should also consider thin or \emph{geometrically vulnerable} structures that may represent a worst-case scenario for thermally activated deformation (i.e., shrinking and warping) mechanisms. 

The choice of solvent has a significant effect on the observed response. In a saltwater solution, the specimen reached mass uptake saturation within a day, whereas in DI water it did not reach saturation over 30 days. The solvent uptake is significantly higher (e.g., 35\% at 95$^\circ$C) in DI water when compared to saltwater (15\% at 95$^\circ$C). This behavior is consistent with the higher net chemical potential of water in DI water, which provides a greater driving force for solvent uptake than in saltwater. In the saltwater, the high concentration of dissolved ions can bind the surrounding water molecules through ion-dipole interaction and effectively lower
the chemical potential for the water, reducing the solvent uptake in the AM polymer. Some key observable differences are shown in Table \ref{table_DIvsSaltWater}.
\begin{table}[h]
    \centering
    \caption{Summary of diffusion characteristics for saltwater and DI water exposures.}\label{table_DIvsSaltWater}
    \begin{tabular}{l|p{5.5cm}|p{5.5cm}}
        \hline
        \vspace{1mm}
        \textbf{Solvent} & \textbf{Saltwater (3.5\%)} & \textbf{DI water} \\
        \hline
        \vspace{1mm}
        Mass saturation & Reaches saturation in 1 day & Continues to rise over 30 days \\
        \hline
         \vspace{1mm}
        Mass uptake (30 days) & 5\% (@65$^\circ$C); 15\% (@95$^\circ$C) & 20\% (@65$^\circ$C);   35\% (@95$^\circ$C)  \\
        \hline
        \vspace{1mm}
        Solvent uptake behavior & Solvent uptake reduced by dissolved ions & Higher chemical potential driving solvent uptake  \\
 \hline
    \end{tabular}
\end{table}
For 50$^\circ$C and below, no qualitative or quantitative changes were observed in either saltwater or DI water. Consequently, we propose a material-dependent temperature limit, found to be 50$^\circ$C for ABS, below which accelerated aging does not activate physical mechanisms beyond those representative of in-service degradation. More broadly, these findings motivate exploring alternative accelerated aging strategies beyond simply increasing temperature, such as modifying salt concentration or hydrostatic pressure, which may accelerate aging while preserving the relevant degradation physics, which will be the focus of our future work.

\section{Conclusion}
We have shown that while temperature is an important parameter for accelerated aging, special care must be taken in its use, as elevated temperatures can activate additional mechanisms that are not present under typical service conditions, thereby altering the physics governing degradation. Consequently, using temperature as the sole acceleration factor, without considering physical limits, may produce degradation mechanisms that are not representative of real in-service behavior. The key contributions and findings from this study are as follows:
\begin{enumerate}

\item We present a systematic method to identify an accelerated-aging temperature threshold for polymers, using AM ABS as a representative material. For the exposure conditions investigated here (air, saltwater, and DI water), the accelerated aging temperature threshold was identified as 50$^\circ$C. The dimensional changes were negligible, and no distortions were observed for exposures at 50$^\circ$C and below. 

\item The results show that exposure at 65$^\circ$C and 95$^\circ$C, although well below the glass transition temperature, induces significant shrinkage along the printing direction and out-of-plane warping, resulting in structural deformations that are not representative of in-service conditions. We further show that elevated temperatures alone can induce these deformations, likely through the relaxation of process-induced internal stresses introduced during FDM 3D printing, resulting in partial recovery toward the material's initial configuration.

\item Although the observed geometric distortions are primarily temperature-driven, their severity can be significantly amplified by the presence of solvents such as saltwater and DI water. These findings highlight that a valid accelerated aging protocol must account for the effects of (i) temperature, (ii) solvent, and (iii) their coupling. 

\item The results show that thermally induced shape distortions can remain undetected in thicker (bulk) cubic specimens, which exhibit negligible deformation, whereas thin structures are far more susceptible to thermally activated shrinkage and warping. These findings highlight the importance of assessing accelerated aging protocols using geometrically vulnerable structures rather than relying solely on thicker (bulk) specimens.

\item Exposure to DI water and saltwater exhibited similar overall trends but differed substantially in magnitude. DI water did not reach mass-uptake saturation over the 30-day exposure period, whereas saltwater reached saturation within approximately one day. In addition, DI water resulted in substantially greater solvent uptake and more pronounced geometric distortions. More importantly, the inclusion of both exposure conditions enabled the effects of solvent uptake and temperature to be decoupled, providing insight into their individual contributions to the observed structural response.
 
\end{enumerate}

Broadly, the present work highlights that accelerated aging methodologies must be designed with careful consideration of polymer-specific physical limits, as indiscriminate increases in temperature may alter the integrity of the structure or underlying degradation physics and lead to behavior that is not representative of service conditions. Based on the results, we propose the existence of a material-dependent temperature limit, below which accelerated aging by increasing temperature remains physically representative and does not activate nonphysical deformation or degradation mechanisms. These findings further motivate exploring alternative acceleration strategies beyond temperature alone, such as modifying salt concentration. The differences in mass uptake and resultant structural responses of AM polymer exposed to different solvents suggest a potential paradigm where changing salt concentration (possibly lowering) can accelerate the hydrolytic seawater aging for marine applications. The present study represents a first step toward establishing physically consistent accelerated-aging protocols with a suggestion to further extend the methodology to other AM processes and materials.



\section*{Acknowledgments}
This research was sponsored by the Department of the Navy, issued by the Office of Naval Research under the Naval Engineering Education Consortium grant award number N00178-25-1-0004. We are also grateful to the Brown School of Engineering for the Hibbitt Postdoctoral Fellowship support to K. Alkhoury.

\section*{Declaration of Interest}
The authors declare no competing interests. 
\clearpage

\bibliographystyle{unsrtnat}
\bibliography{References}

\end{document}